\documentclass[
reprint,amsmath,amssymb,prl,aps]{revtex4-2}

\usepackage{graphicx}
\usepackage{dcolumn}
\usepackage{bm}
\usepackage{float}
\usepackage[usenames,dvipsnames]{color}

\begin{document}

\preprint{APS/123-QED}

\title{$T$-square electric resistivity and its thermal counterpart in RuO$_2$}

\author{Yu Ling$^{1,2}$, Florent Pawula$^{3,4}$  Ramzy Daou$^{3}$, Benoît Fauqué$^1$ and Kamran Behnia$^1$}

\affiliation{$^1$Laboratoire de Physique et d'Etude des Matériaux (CNRS) ESPCI Paris, PSL Universit\'e, 75005 Paris, France \\
$^2$Wuhan National High Magnetic Field Center and School of Physics, Huazhong University of Science and Technology, Wuhan 430074, China \\
$^3$Laboratoire de Cristallographie et Sciences des Mat\'eriaux,
Universit\'e de Caen Normandie, ENSICAEN, CNRS, Normandie Univ, CRISMAT UMR6508, F-14000 Caen, France\\
$^4$Nantes Universit\'e, CNRS, Institut des Mat\'eriaux de Nantes Jean Rouxel (IMN)-UMR 6502, 44000 Nantes, France}

\begin{abstract}
 We present a study of low-temperature electric and thermal transport in RuO$_2$, a metallic oxide which has attracted much recent attention. Careful scrutiny of electric resistivity reveals a quadratic temperature dependence below $\sim$ 20 K  undetected in previous studies of electronic transport in this material. The prefactor of this T$^2$ resistivity, given the electronic specific heat, corresponds to what is expected by the Kadowaki-Woods scaling. The variation of its amplitude across 4 different samples is negligible despite an eightfold variation of residual resistivity. There is also a T$^5$ resistivity due to scattering by phonons. By measuring thermal conductivity, $\kappa$, at zero field and at 12 T, we separated its electronic and phononic components and found that the former respects the Wiedemann-Franz law at zero temperature and deviates downward at finite temperature. The latter corresponds to a threefold discrepancy between the prefactors of the two (thermal and electric) T-square resistivities. Our results, establishing RuO$_2$ as a weakly correlated Fermi liquid, provide new input for the ongoing theoretical attempt to give a quantitative account of electron-electron scattering in metallic oxides starting from first principles. 
\end{abstract}

\maketitle

RuO$_2$ \cite{Schafer1963} crystallizes in the tetragonal rutile structure with space group symmetry of P4$_2$/mnm($D_{4h}^{14}$). It belongs to a family of MO$_2$ rutile compounds (M=Ru, Ir, Os, Cr, Re, Mo, W is a transition metal) in which the building blocks are distorted octahedra of MO$_6$. For many decades, RuO$_2$ was known as a metallic oxide \cite{1969electrical} and a Pauli paramagnet \cite{Ryden1970}. Its metallic grains embedded in a glassy matrix become a popular thick-film resistive thermometer in cryogenic environments \cite{BATKO19921167,Ptak2004}. Crystalline RuO$_2$ has attracted recent interest following reports of antiferromagnetic ordering above room temperature, leading to its identification as an altermagnet \cite{altermagnet}. This conjecture was refuted by more recent studies concluding that any magnetic ordering is either absent or the magnetic moment undetectably small \cite{Hiraishi2024,Kiefer_2025,YUMNAM2025102852}. 

This metallic oxide has a relatively low room-temperature resistivity and its residual resistivity ratio can become remarkably large \cite{MARCUS1968518}.  Several experimental \cite{SLIVKA1968169,Magnetothermaloscilla} and theoretical \cite{1976electronic,Abinitio,local2021} studies have explored its Fermi surface. Investigations of transport properties \cite{1969electrical,1976electronic,1993Electrontransport,Multiband,universalscaling} have found that it is a compensated metal \cite{universalscaling}, has  multiple bands \cite{Multiband} and conforms to the Bloch-Gr\"uneisen picture of resistivity \cite{1976electronic,universalscaling}. However, in contrast to  other metallic oxides \cite{Tokura1993,Maeno1997,Inoue1998,Grigera2001,Nakamae2003,STOlin2015,Tomioka2021}, there has been no report of T-square resistivity, which is expected and observed in metallic Fermi liquids \cite{Pal2012,behnia2022origin}. 

The phase space for electron-electron scattering grows quadratically with temperature. It has been known for decades \cite{Landau1936,Baber1937} that this leads to a $T^2$ contribution to the electric resistivity, $\rho$, in addition to the $T^5$ term originating from electron-phonon scattering \cite{ziman2001electrons}. Therefore, the low-temperature resistivity is expected to follow:
\begin{equation}
\rho=\rho_0+A_2T^2+A_5T^5
\label{equ1}
\end{equation} 
Here, $\rho_0$ is the residual resistivity due to disorder.  $A_2$ and $A_5$ are the prefactors of the $T^2$ and $T^5$ resistivity terms, respectively. Their relative weight determines the temperature window of their prominence. 

The microscopic mechanism by which these collisions degrade conduction has not been
unambiguously identified. In dilute metals, such as n-doped strontium titanate \cite{STOlin2015}, 
Bi$_2$O$_2$Se \cite{BiOSe2020t} (but also ZrTe$_5$ \cite{Santos2020} and Bi$_2$Se$_3$\cite{Singha2024}), T-square resistivity persists despite the fact that the Fermi surface is too small for Umklapp events. In all these cases, there is a single Fermi surface at the center of the Brillouin zone. The absence of the two known routes from fermion-fermion collisions to dissipation in these cases, has revived a debate on the origin of the T-square resistivity in Fermi liquids \cite{Kumar2021,behnia2022origin,Kumar2024,behnia2024heat}.

Despite this uncertainty about its origin, the order of magnitude of A$_2$ in a Fermi liquid, given the knowledge of its other experimentally measurable properties, can be guessed. In dense metals, $A_2$ scales with the square of the electronic specific heat \cite{kadowaki1986heavy,Hussey2005}. In dilute metals, where the specific heat depends not only on the Fermi energy but also on the carrier density, the relevant scaling is between $A_2$ and the inverse of the square of the Fermi energy \cite{BiOSe2020t,behnia2022origin}. These are known as [extended] Kadowaki-Woods (KW) scaling.
\begin{figure*}
    \centering
    \includegraphics[width=0.9\textwidth]{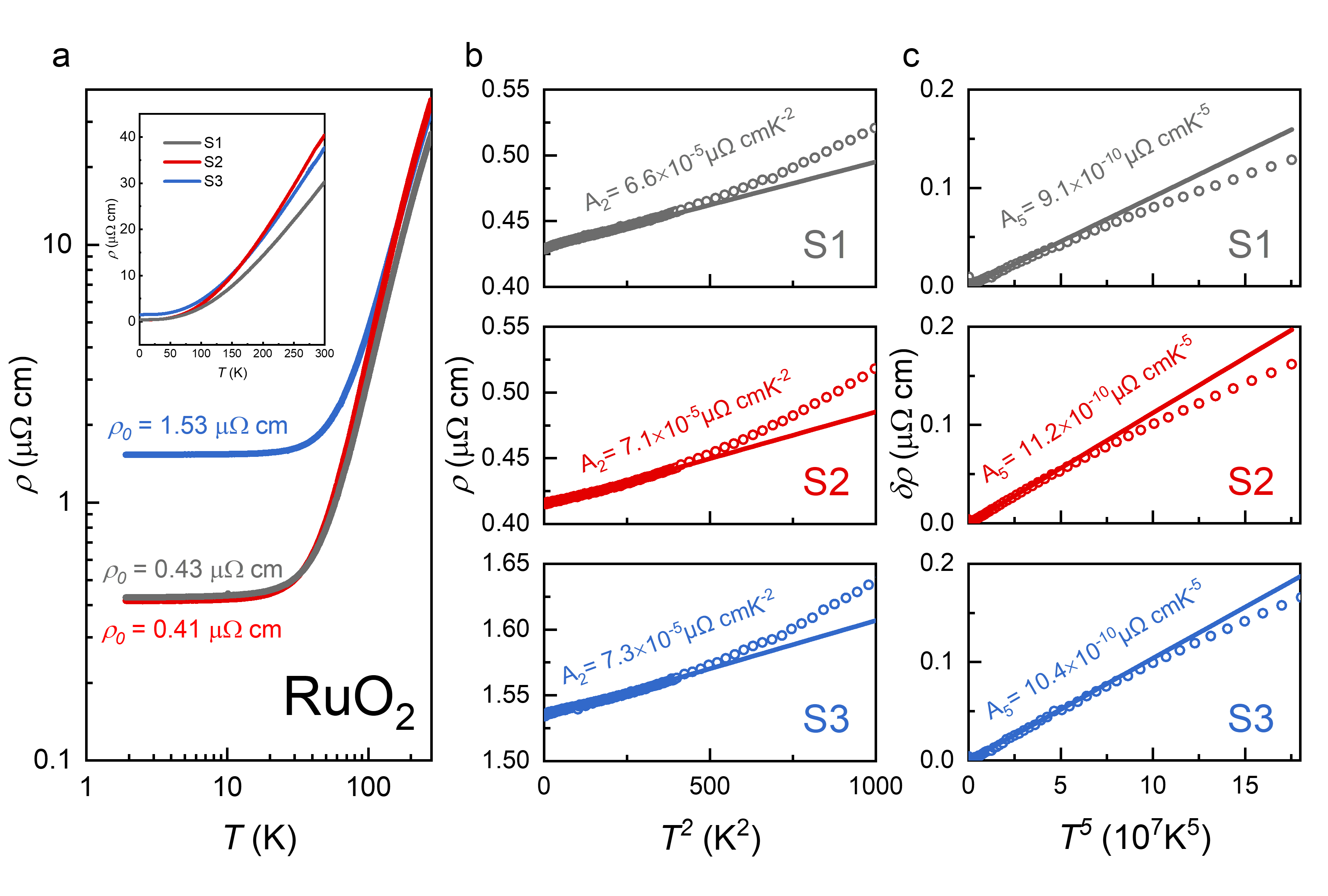}
    \caption{\textbf{Temperature-dependent resistivity of RuO$_2$. } (\textbf{a}) Resistivity plotted on a log-log scale over the full temperature range. The inset shows a linear scale. (\textbf{b}) Resistivity versus $T^2$ for samples S1-S3 below 30 K. Experimental data (open circles) can be fit to $\rho=\rho_0+A_2T^2$ (solid lines) below $\approx$ 20  K. An upward deviation indicates the predominance of a larger exponent at higher temperatures. (\textbf{c}) The subtracted resistivity $\delta\rho=\rho-\rho_0-A_2T^2$ plotted versus $T^5$ for samples S1-S3. Solid lines represent fits to $\delta \rho=A_5T^5$ below $\approx$ 40 K. A downward deviation is visible at higher temperatures. }
    \label{fig:1}
\end{figure*}

Quantifying the amplitude of the contribution of electron-electron (\textit{e-e}) scattering to $\rho$ in presence of correlations has emerged as a new challenge to computational condensed-matter physics. In two metallic oxides (SrVO$_3$ and SrMoO$_3$), it was computed from first principles by a combination of single-site dynamical mean-field theory and density functional theory \cite{kugler2024calcu}. Comparison with experiment was hindered, however, by the surprising inconsistency between the data reported by different groups investigating crystals and thin films.

Here,  we report on a study of low-temperature transport in  RuO$_2$ samples with residual resistivity ratios (RRR) varying between 12 and 99. A careful study of c-axis resistivity led to the quantification of $A_2$ and $A_5$ in all samples. The detection of the T-square term required a careful scrutiny of the data below 20 K. We find that the amplitude of this term is in agreement with what is expected by KW scaling.  We then carried out thermal conductivity measurements on the cleanest sample. To disentangle the phononic and electronic contributions to heat transport, we repeated the experiments in a  magnetic field of 12 T. The electronic thermal resistivity was found to verify the Wiedemann-Franz law at low temperature and exhibit a quadratic temperature dependence as previously found in other compensated metals. A quantitative account of our findings from first principles emerges as a challenge for computational condensed matter physics.  
\begin{figure*}
\centering
    \includegraphics[width=0.95\textwidth]{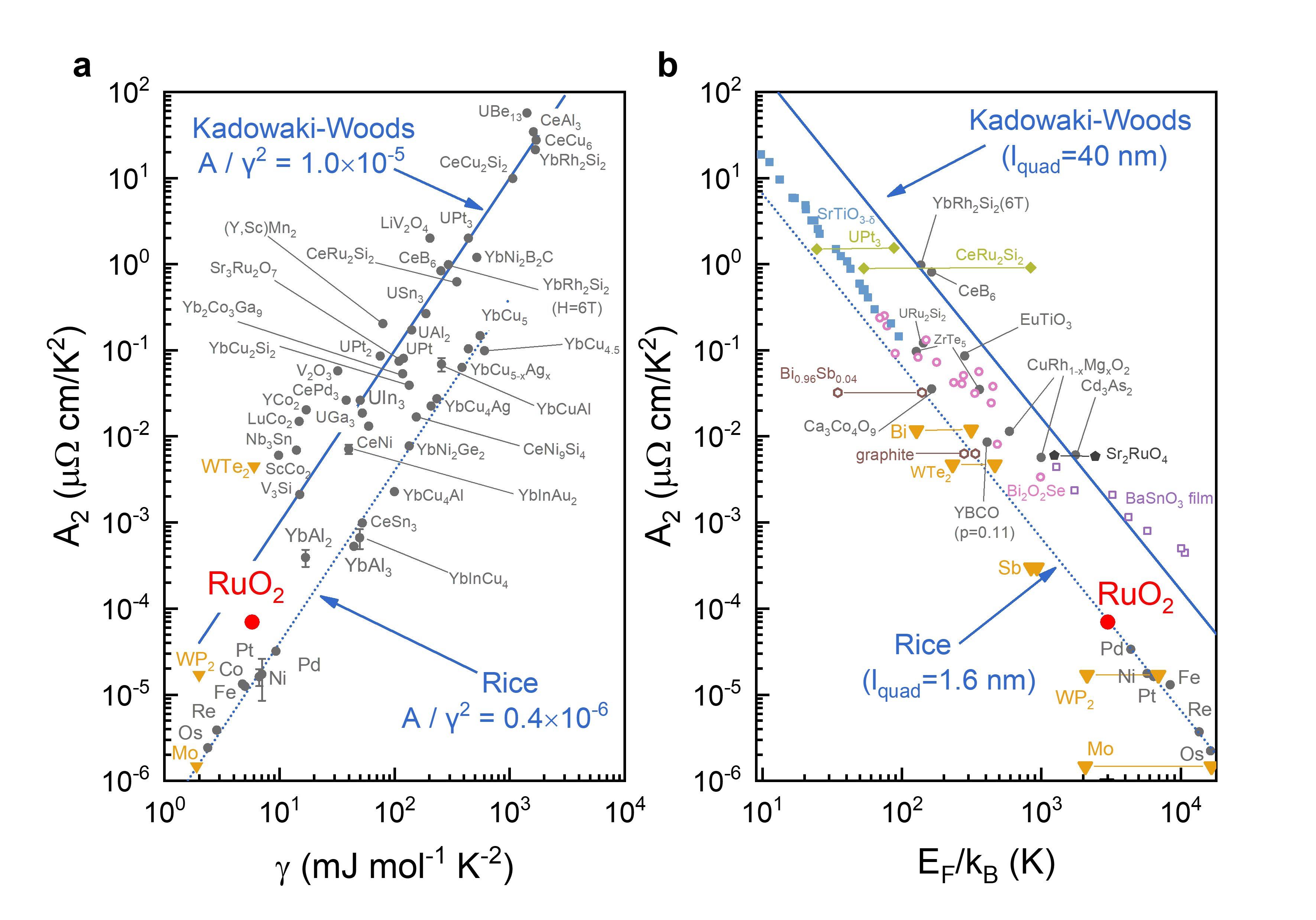}
    \caption{\textbf{RuO$_2$ in Kadowaki-Woods plots.} (\textbf{a}) The prefactor A$_2$ plotted as a function of fermionic specific heat $\gamma$ on a log-log scale. RuO$_2$: red circle (this work); data collected from \cite{tsujii2003}: gray circle; semimetal WTe$_2$, WP$_2$ and Mo\cite{gourgout2024electronic}: orange inverted triangle. (\textbf{b}) The prefactor A$_2$ plotted as a function of Fermi temperature $E_F/k_B$ on a log-log scale. RuO$_2$: red circle(this work); data collected from \cite{BiOSe2020t}: gray circle; BaSnO$_3$ film: purple open square; Bi$_2$O$_2$Se: pink open circle; Sr$_2$RuO$_4$: black pentagon; UPt$_3$, CeRu$_2$Si$_2$: green diamond; Bi$_{0.96}$Sb$_{0.04}$, graphite: brown open hexagon; SrTiO$_{3-\delta}$\cite{STOlin2015}: blue square; semimetal Bi, Sb, Mo, WTe$_2$, WP$_2$\cite{gourgout2024electronic}: orange inverted triangle. Most materials are located between the upper (Kadowaki-Woods) bound and the lower (Rice) bound (blue dotted line). }
    \label{fig:2}
\end{figure*}

 \begin{table*}
\caption{\label{tab:table1}\textbf{Characteristics of the RuO$_2$ samples used in this study.} Residual Resistivity Ratio ($RRR$) defined as $\rho(300K)/{\rho_0}$. The carrier mean free path $l_0$ is derived from Drude model using the residual resistivity and Fermi wave vector $k_F$ . The transport mobility is $\mu_0=1/\rho_0e(n_e+n_h)$, with the carrier concentration $\bar{n}=n_e=n_h=8.87\times 10^{21}cm^{-3}$\cite{universalscaling}. The units of $A_2$ and $A_5$ are $10^{-5}\mu \Omega$ $cm K^{-2}$ and $10^{-10}\mu \Omega$ $cm K^{-5}$ respectively.}
\begin{ruledtabular}
\begin{tabular}{cccccccc}
 sample &Size ($mm^3$) &RRR &$\rho_0$($\mu \Omega$ $cm$) & $A_2$ &$A_5$ &l$_0$($\mu m$)&$\mu_0$($m^2V^{-1}s^{-1}$) \\ \hline
S1&1.2$\times$0.2$\times$0.067&70&0.43&6.6&9.1&0.47&0.081\\ 
S2&0.78$\times$0.18$\times$0.068&99 &0.41&7.1&11.2&0.50&0.085\\
S3&0.94$\times$0.29$\times$0.066&25&1.53&7.3&10.4&0.13&0.023\\
S4&1.0 $\times$ 0.2$\times$0.1 &12&3.26&7.7&11.7&0.06&0.011\\
\end{tabular}
\end{ruledtabular}
\end{table*}

Single crystals of RuO$_2$ were grown using vapor-transport technique in a multi-zone tubular furnace \cite{HUANG19821305}. A mullite tube was placed in the furnace operating under flowing O$_2$. An alumina crucible containing 5 g of commercially available RuO$_2$ powder (Chempur, 99.9\%) was introduced in the first third of the tube length. The O$_2$ flow was initially set to 2 cm$^3$/min while the temperature at the furnace center was increased up to 1350$^\circ$ C over 24 h, and subsequently ramped up to 60 cm$^3$/min for the following 12 days. The furnace was then powered down and allowed to cool overnight, after which the O$_2$ flow was turned off eventually. The resulting single crystals are elongated in shape, with typical lengths of 0.5-1 mm. The XRPD patterns of the commercial starting powder and the ground single crystals were found to be very similar. The single crystals are single-phase rutile RuO$_2$. Electron diffraction analysis confirms the space group and lattice parameters. Lattice parameters obtained from Le Bail refinements using a tetragonal model with space group P4$_2$/mnm (No. 136) are a = 4.4908(2) \AA~ and c = 3.1063(1) \AA~ for the single crystals, in excellent agreement with previous reports \cite{BUTLER197181} (See the supplementary Materials for more details on sample characterization).

Electrical resistivity of RuO$_2$ samples was measured by a standard four-wire method with 25 $\mu m$ diameter gold wires connected with silver paste to the sample. Below 30 K, each measurement was performed after a stabilization time of 2 minutes in order to improve the resolution. Data for three samples are shown in Fig.\ref{fig:1} a. The room temperature resistivity values of three samples fall within the range of 30-40 $\mu \Omega$ $cm$, consistent with earlier reports \cite{1969electrical,1993Electrontransport,Multiband,universalscaling}. The  behavior below 30 K is highlighted in Fig.\ref{fig:1} b. One can see that resistivity follows a $T^2$ dependence below $\approx$ 20 K. The upward deviation signals the presence of another larger exponent for inelastic resistivity, due to electron-phonon scattering.

To extract the $A_5$ prefactor of the $T^5$ dependence, we subtracted the impurity and \textit{e-e} scattering terms from the total resistivity and plotted the remainder as a function of T$^5$. As illustrated in Fig.\ref{fig:1} c, the slope below 40 K yields $A_5$. In each sample, it was found to be $\approx 1\times10^{-9}$ $\mu \Omega$ $cmK^{-5}$. Above 40 K, the resistivity exhibits a clear downward deviation from the $T^5$ trend. This is consistent with what is expected in the Bloch-Gr\"uneisen picture, in which resistivity is $\propto T^5$ at temperatures below $\Theta_D/10$ ($\Theta_D$ is the Debye temperature) and becomes linear at higher temperatures as electron-phonon scattering becomes elastic. Heat capacity measurements of RuO$_2$ \cite{969heatcapacity} indicate a $\Theta_D$ exceeding 610 K. 

The properties of our RuO$_2$ samples are summarized in Table \ref{tab:table1}. Note that while the residual resistivity ratio (RRR) of the cleanest sample is eight times larger than that of the  dirtiest one, the extracted $A_2$ and $A_5$ differ by a modest factor of 1.2. Such a difference is not large enough to be considered significant.  After this paper was submitted, Paul and co-workers reported on transport and thermodynamic measurements on ultra-clean \cite{PAUL2026128405} RuO$_2$ single crystals with a RRR of 1200 and 400 \cite{paul2025nonanalyticfermiliquidcorrectionspecific} and confirmed the  T$^2$  temperature dependence of the electric resistivity. Their reported T-square prefactors (A$_2= 5.3-30 \times 10^{-5}\mu \Omega$ $cm K^{-2}$) is to be compared with ours  (A$_2= 6.6-7.7 \times 10^{-5}\mu \Omega$ $cm K^{-2}$). The order of magnitude  of A$_2$ is consistent between the two studies.

\begin{figure*}
\centering
    \includegraphics[width=0.8\textwidth]{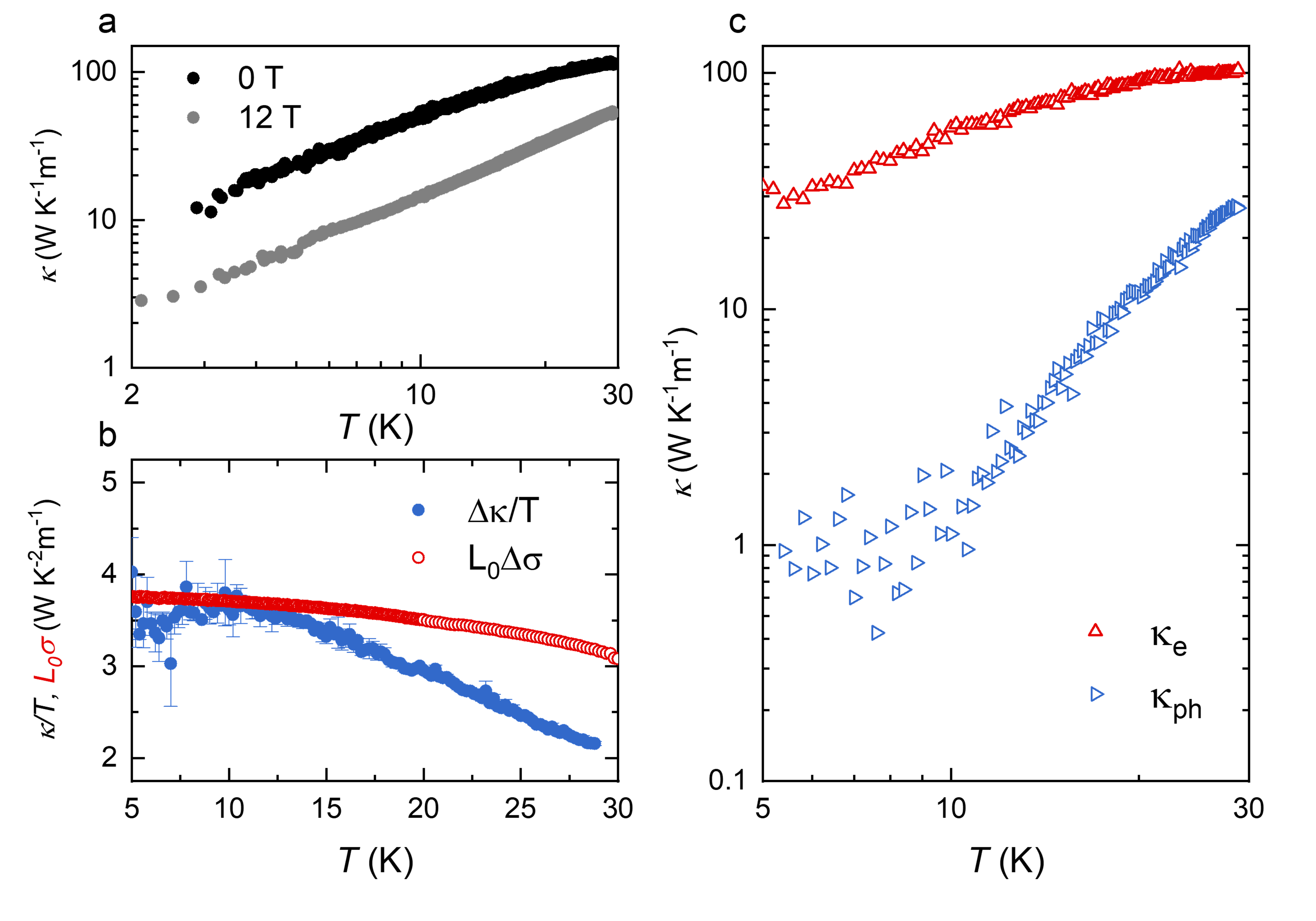}
    \caption{\textbf{Temperature-dependent thermal conductivity of RuO$_2$ below 30 K.} \textbf{a} Temperature-dependent thermal conductivity of RuO$_2$ plotted on log-log scale in 0 T (black circle) and 12 T (gray circle) magnetic fields. \textbf{b} Temperature-dependence of $\Delta \kappa/T$ (blue circle) and $L_0\Delta \sigma$ (red open circle). $\Delta \kappa $ and $\Delta \sigma$ equal to $\kappa(0 T)-\kappa(12T)$ and $\sigma(0T)-\sigma(12T)$ respectively. Error bars caused by signal fluctuations and geometrical errors are shown in the figure. \textbf{c} Temperature-dependence of electronic (red triangle) and phononic (blue triangle) contributions to thermal conductivity in a log-log scale.}
    \label{fig:3}
\end{figure*}

Scaling between the amplitude of $A_2$ and the electronic specific heat coefficient $\gamma$ was noticed by Rice \cite{rice1968electron} for elemental transition metals and  by Kadowaki and Woods for heavy-fermion systems \cite{kadowaki1986heavy}. On a log-log scale, there is a correlation between two measurable quantities, which both depend on the density of states. It holds in a wide variety of Fermi liquids, despite the fact that the microscopic origin of dissipation associated with \textit{e-e} scattering is yet to be identified (Fig.\ref{fig:2} a).  However, in dilute metals such as bismuth and antimony, the extremely low carrier concentrations (in the range of $\sim 10^{17}$ to $10^{19}\mathrm{cm^{-3}}$; corresponding to one mobile electron per thousands of atoms), do not fit to this scaling. This deviation arises because the electronic specific heat coefficient $\gamma$ depends on the carrier density, whereas the prefactor $A_2$, mainly set by the degeneracy temperature of the fermionic system, does not \cite{STOlin2015,BiOSe2020t,behnia2022origin}. These dilute metals fit to an extended scaling plot where the Fermi temperature $E_F/k_B$ replaces $\gamma$ plot (Fig.\ref{fig:2} b).

To locate RuO$_2$ on the Kadowaki--Woods plot, one needs to know the Sommerfeld coefficient, in other words, the electronic specific heat, $\gamma$. This is a thermodynamic property and is not expected to correlate with disorder and to be sample dependent. The T-square prefactor $A_2$ obtained from our measurements combined with $\gamma$ ($5.77~\mathrm{mJ~mol^{-1}~K^{-2}}$ reported by Passenheim and McCollum \cite{969heatcapacity} and  $5.07~\mathrm{mJ~mol^{-1}~K^{-2}}$ according to Paul \textit{et al.} \cite{paul2025nonanalyticfermiliquidcorrectionspecific}) gives rise to the red circle in Fig.\ref{fig:2} a. The Fermi temperature of RuO$_2$ can be estimated to be $\approx 3000~\mathrm{K}$, based on the slope of the Seebeck coefficient reported in \cite{Multiband} and its empirical relationship with $E_F$  \cite{behnia2004thermoelectricity,kurita2019correlation}. As shown in Fig.\ref{fig:2} b, RuO$_2$ conforms to the extended version of the Kadowaki--Woods scaling too.  

Let us also recall that RuO$_2$ conforms to another universal relation linking two experimentally accessible properties of a Fermi liquid. Decades ago, Ryden and Lawson \cite{Ryden1970}  noticed that the ratio of the electronic specific heat and the magnetic susceptibility in RuO$_2$ conforms to the Wilson ratio linking the Pauli susceptibility and the Sommerfeld coefficient \cite{solyom2008fundamentals}. This is another confirmation that this solid is a Pauli paramagnet with no magnetic ordering.  

We now turn our attention to thermal transport. Since thermal resistivity arising from \textit{e-e} scattering does not require Umklapp scattering, its origin is expected to be more straightforward \cite{macdonald1980electron,behnia2022origin,gourgout2024electronic}. Experimentally, however, obtaining reliable thermal conductivity data is  more challenging than measuring electrical conductivity. We measured the thermal conductivity of the RuO$_2$ sample with the largest RRR (sample S2) using a standard one-heater–two-thermometers method. Fig.\ref{fig:3} a presents the temperature-dependence of thermal conductivity $\kappa$ of RuO$_2$ below $30~\mathrm{K}$ at zero field as well as in a magnetic field of $12~\mathrm{T}$. Our zero-field data is consistent with earlier reports \cite{millstein1970thermal,gladun1980ruo2thermal}. 

Assuming that the phononic contribution to the thermal conductivity is insensitive to the magnetic field, one can separate it from the electronic component upon the application of a magnetic field and by measuring the magnetoresistance of the system. This procedure was successfully employed in several cases in which electrons and phonons both contribute substantially to heat transport \cite{White01041958,STO2023t,gourgout2024electronic,xie2024purity}.

As shown in Fig.\ref{fig:3} a, the application of magnetic field reduces $\kappa$. Given the presence of a substantial electronic component and its magnetoresistance, this is not surprising. Fig.\ref{fig:3} b, compares the field-induced reduction of thermal conductivity (divided by temperature) $\Delta\kappa/T$ with the field-induced reduction of electric conductivity (multiplied by $L_0$), $L_0\Delta\sigma$. Here, $L_0=\frac{\pi^2}{3}\frac{k_B^2}{e^2}$ is the  Sommerfeld value)
\begin{figure}
\centering
    \includegraphics[width=1.1\columnwidth]{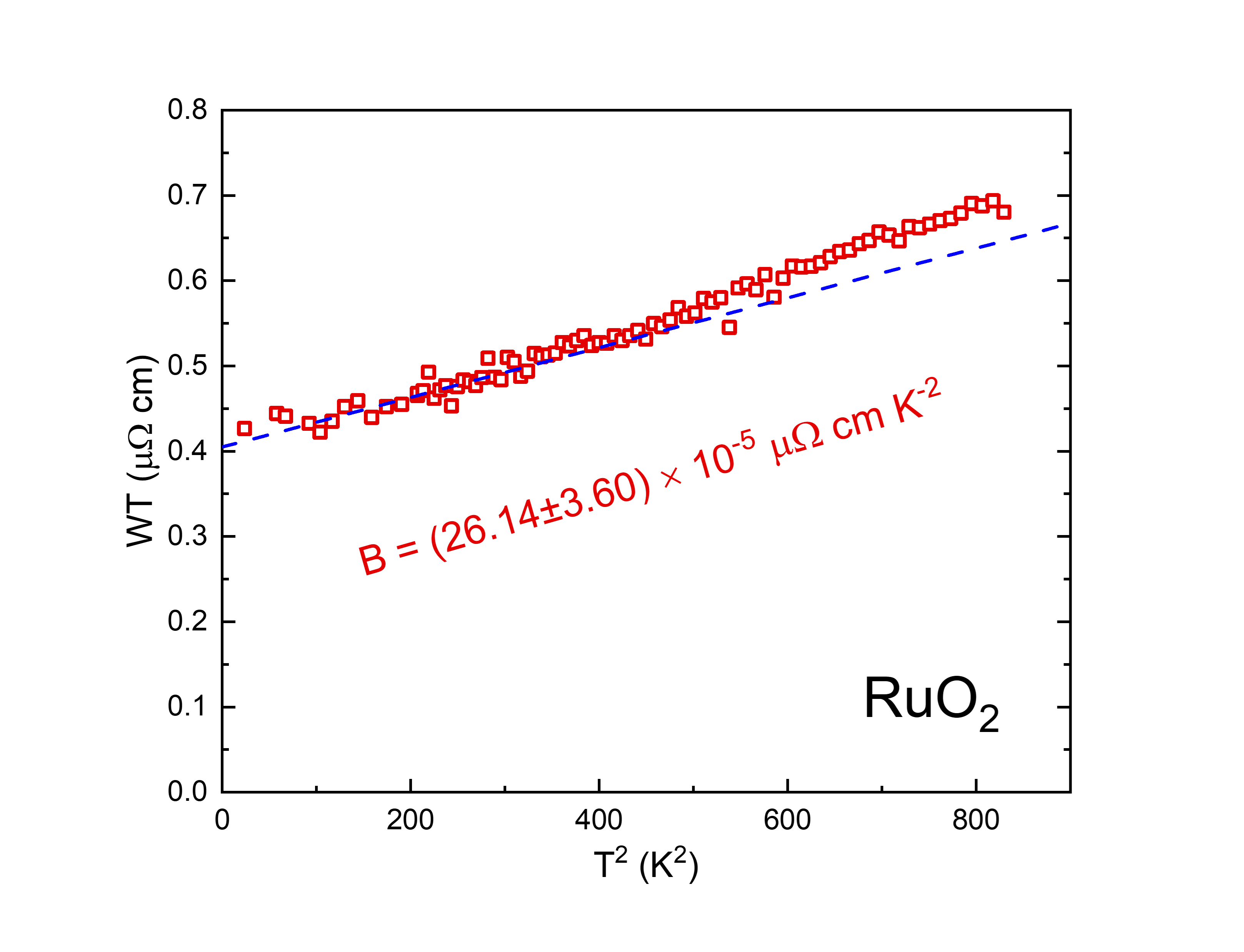}
    \caption{\textbf{Temperature dependence of electronic thermal resistivity} $WT$ is defined as the inverse of $\kappa/T$.  Normalized by $L_0$, it can be expressed in units of electric resistivity. Plotting it as function of T$^2$ reveals an intercept, due to scattering by impurities and a slope, due to \textit{e-e} scattering.}
    \label{fig:4}
\end{figure}
\begin{table*}
\caption{\label{tab:table2}\textbf{ T-square resistivities in semimetals}. Carrier density, $A_2$,  $B_2$ and their ratio for semimetals in which $B_2$ has been quantified. }
\begin{ruledtabular}
\begin{tabular}{cccccc}
System &\textit{n = p} $(cm^{-3})$&$A_2$ ($n\Omega$ cm/$K^2$) &$B_2$ ($n\Omega$ cm/$K^2$)&$B_2/A_2$ & Refs.\\
\hline
Bi&3.0 $\times$ $10^{17}$ & 12&35&2.9&\cite{gourgout2024electronic}\\ 
Sb&5.5 $\times$ $10^{19}$& 0.3&0.6&2.0&\cite{Sb2021thermal} \\
WTe$_2$&6.8 $\times$ $10^{19}$& 4.5&11&2.4&\cite{xie2024purity}\\
WP$_2$&2.5 $\times$ $10^{21}$& 0.017&0.074&4.6&\cite{WP2018departure}\\
W&2.5 $\times$ $10^{22}$& 8.7 $\times$ 10$^{-4}$&6 $\times $ 10$^{-3}$&6.9&\cite{desai1984electrical,Wagner1971}\\
RuO$_2$&8.8 $\times$ $10^{21}$& 0.071&0.26&3.7&This work
\end{tabular}
\end{ruledtabular}
\end{table*}
The convergence of these two quantities at low temperature confirms the validity of the Wiedemann-Franz law. At higher temperatures ($T > 10$~K), a downward deviation emerges, as reported in other metals \cite{WP2018departure,Sb2021thermal,gourgout2024electronic,xie2024purity,CeRhIn5thermal,1994UPt3thermal}. It is a consequence of the distinction between horizontal and vertical scattering events in relaxing momentum. The combination of 0~T and 12~T data allows the extraction of the electronic thermal conductivity $\kappa_e$ assuming \cite{STO2023t}:
\begin{equation}
    \kappa_e(\mu_0H=0)=\Delta\kappa\frac{\sigma(\mu_0H=0)}{\Delta\sigma}
    \label{equ3}
\end{equation}
In other words, we are neglecting any variation of the Lorenz number between zero field and 12 T at a given temperature. The phonon contribution is then obtained as $\kappa_{\mathrm{ph}} = \kappa - \kappa_e$. 

As shown in Fig.\ref{fig:3} c, $\kappa_e$ exceeds $\kappa_{\mathrm{ph}}$ by more than one order of magnitude. Given the large Fermi surface of RuO$_2$ and the reasonably long mean free path of charge carriers, the domination of the electronic component is not surprising.

Defining the electronic thermal resistivity  as $WT = \frac{ T}{\kappa_e}$ (and multiplying it by $L_0$ in order to express it in units of electrical resistivity) allows a quantitative comparison of heat and charge transport by electrons. In a Fermi liquid, analogously to  electrical resistivity, $WT$ follows a quadratic temperature dependence at low temperatures:
\begin{equation}
    WT=(WT)_0+B_2T^2
    \label{equ4}
\end{equation}
where $(WT)_0$ is the residual term associated with impurity scattering, and $B_2$ is the prefactor of the $T^2$ term in thermal resistivity. Fig.\ref{fig:4} displays $WT$ of RuO$_2$ plotted as a function of $T^2$, revealing a low-temperature linear dependence from which the prefactor $B_2$ is extracted. It is more than three times larger than $A_2$. As seen in Table  II, which lists the reported $B_2/A_2$ ratio in various compensated metals, the case of RuO$_2$ does not stand out. The  ratio varies between 2 and 7.  This is in contradiction with early  theoretical works suggesting an upper boundary to this ratio \cite{herring1967simple,bennett1969exact}. However, more recent theoretical works \cite{Li2018,Levchenko2025,takahashi2025} do not confirm the existence of this upper limit. 

In addition to semimetals \cite{WP2018departure,Sb2021thermal,xie2024purity,gourgout2024electronic}, the prefactor of $T^2$ thermal resistivity has been quantified in heavy-fermion compounds \cite{1994UPt3thermal,CeRhIn5thermal}. Moreover, the thermal conductivity of liquid $^3$He \cite{He1984thermal} is inversely proportional to temperature below $\approx 20$ mK. This temperature dependence is equivalent to a T-square $WT$. Its  prefactor follows the scaling of the metallic Fermi liquids \cite{behnia2022origin}. Note that at higher temperatures, the quasi-particle heat transport in liquid $^3$He is overwhelmed by the contribution of a collective mode \cite{behnia2024heat}. 

The combination of our quantification of phonon thermal conductivity and the reported $T^3$ specific heat \cite{969heatcapacity} leads to the quantification of the phonon mean free path. At 5 K, it remains two orders of magnitude smaller than the sample thickness and far from the ballistic limit. The most plausible explanation of this feature is the scattering of phonons by mobile electrons. 

In summary, we measured low-temperature electrical and thermal conductivities of several RuO$_2$ crystals and extracted the prefactors of $T^2$ and $T^5$ of electrical resistivity. Across four samples, in which the residual resistivity differs by a factor of 8, the measured T-square prefactor was found to match within twenty percent. This indicates that the T-square resistivity is an intrinsic property of RuO$_2$  and in principle computable from first principles. Employing a magnetic field and exploiting the fact that it affects the electronic thermal conductivity, but not the lattice thermal conductivity, we disentangled the electronic and phononic components of the thermal conductivity. We found that the electronic thermal resitivity displays a quadratic behavior as a function of temperature with a prefactor 3.7 times larger than the electric T-square resistivity. Since thermal resistivity caused by e-e collisions does not require these events to be Umklapp, its computation from first principles \cite{kugler2024calcu} is more straightforward, at least, in principle. Our results confirm that RuO$_2$ is a dense weakly correlated metallic Fermi liquid.

This work was supported by a Cai Yuanpei Franco-Chinese cooperation program (No. 51258NK). We also acknowledge a grant from the \^Ile de France regional council.

\bibliography{apssamp}
\clearpage

\begin{center}{\large\bf Supplementary Materials for ``T-square electric resistivity and its thermal counterpart in RuO$_2$"}\\
\end{center}

\renewcommand{\thesection}{S\arabic{section}}
\renewcommand{\thetable}{S\arabic{table}}
\renewcommand{\thefigure}{S\arabic{figure}}
\renewcommand{\theequation}{S\arabic{equation}}

\setcounter{section}{0}
\setcounter{figure}{0}
\setcounter{table}{0}
\setcounter{equation}{0}

\section{Sample characterization}

\begin{figure*}
\centering
    \includegraphics[width=0.9\textwidth]{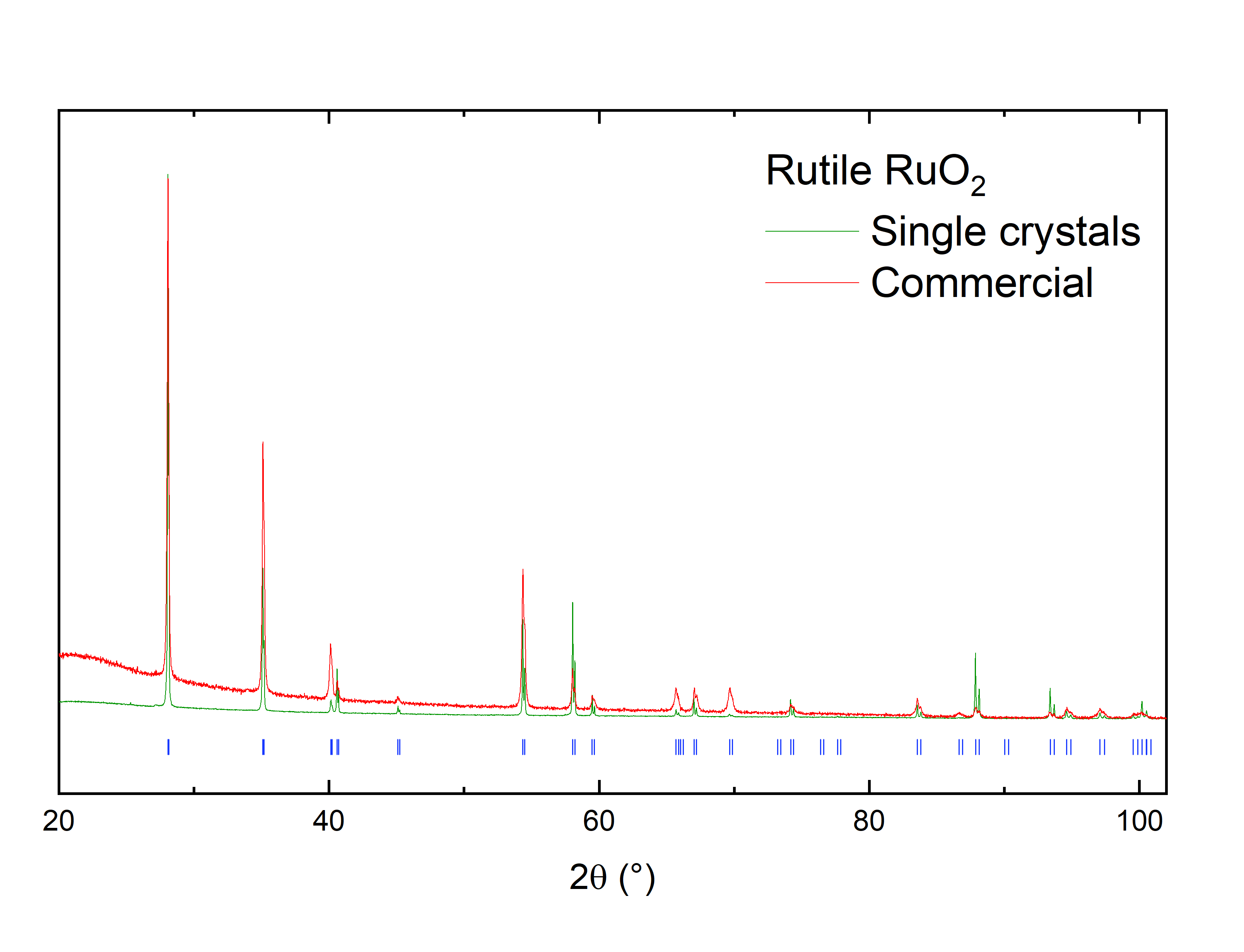}
    \caption{\textbf{XRD characterization} \textbf{a} Powder XRD patterns of crushed single crystals (green) and commercial powder (red) along with the Bragg positions (blue ticks) of a tetragonal model, space group \#136 P42/mnm.}
    \label{fig:XRD}
\end{figure*}

The RuO$_2$ samples were characterized by X-Ray Diffraction (XRD) and by transmission electron microscopy. 
Figure \ref{fig:XRD} shows the powder refraction pattern of crushed single crystals of RuO$_2$. 

\begin{figure*}
\centering
    \includegraphics[width=0.9\textwidth]{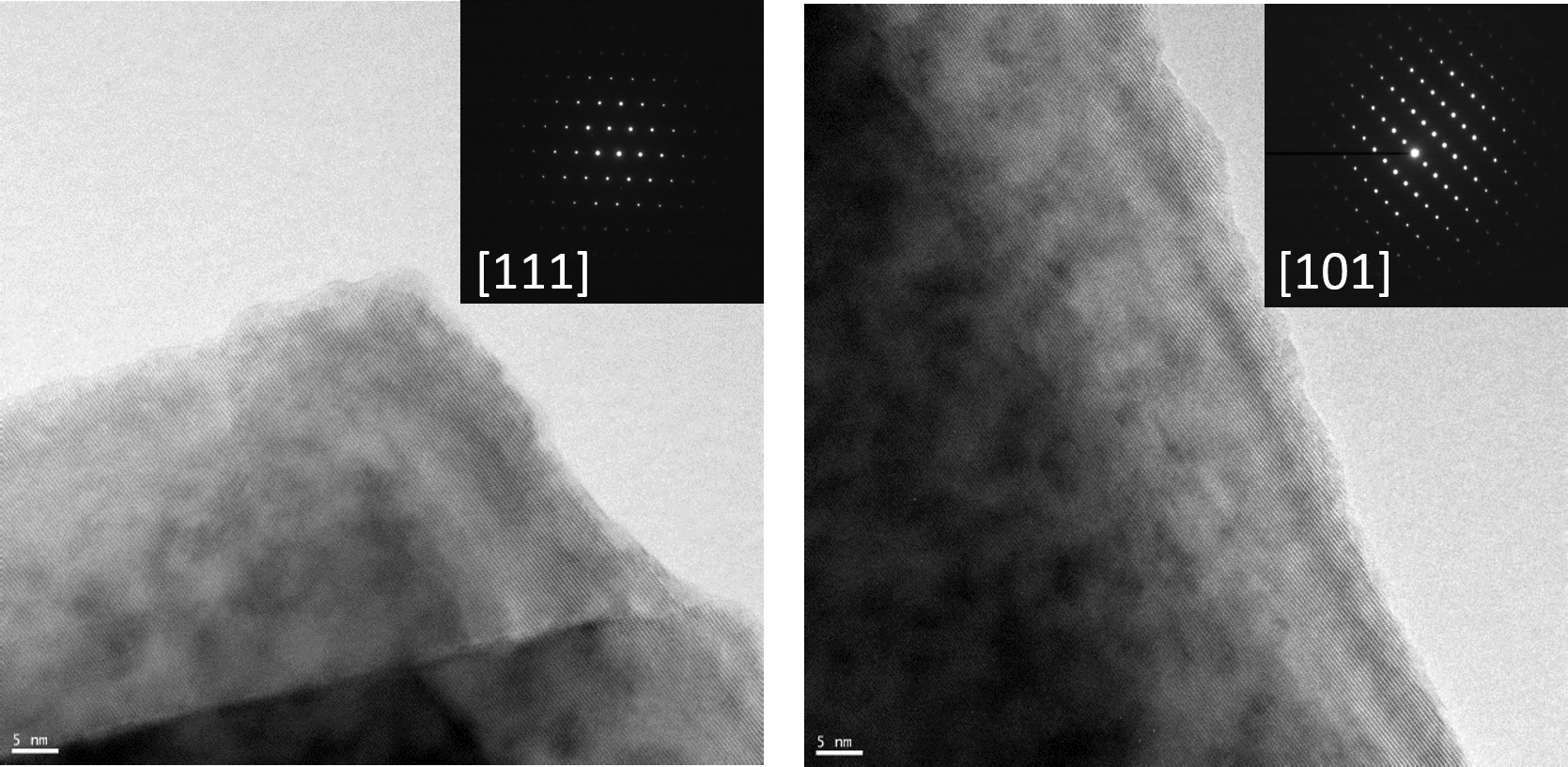}
    \caption{\textbf{TEM Characterization} High resolution TEM images of a single crystal along with ED patterns of the zone axis [111] (left) and [101] (right). Together with the PXRD, this evidences the single crystals are single phase rutile RuO$_2$, and the ED pattern analysis confirms the space group and lattice parameters along the [111] and the [101] directions.}
    \label{fig:TEM}
\end{figure*}

We also made Transmission Electron Microscopy (TEM) images and performed Electronic Diffraction (ED) patterns using two transmission electron microscopes: a FEI TECNAI 30UT (Cs = 0.7 mm) working at 300 kV and a JEOL ARM200 cold FEG double-corrected. Fig \ref{fig:TEM} shows a typical image.


\end{document}